\documentclass[FBSedit,FBSmath,ecsub]{FBSsuppl}
\usepackage{amsfonts}
\usepackage{amssymb}
\usepackage{epsfig}

\title{Photo and Electrodisintegration of Few Body Systems Revisited}
\author{J.-M. Laget\thanks{\textit{E-mail address:} 
jlaget@cea.fr}}
\institute{CEA/Saclay, DAPNIA-SPhN, F91191 Gif-sur-Yvette, France}

\begin{document}

\maketitle
\begin{abstract}
Twenty years after P. Sauer released the state of the art Faddeev solution of
the bound state three nucleon systems, I revisit photo and electrodisengration 
of few body systems with a special emphasis on the prospects opened at Jefferson
Laboratory.
\end{abstract}

\section{Introduction}

In the 80' P. Sauer, and his group at the University of Hanover, solved the Faddeev
equations for the three nucleon systems~\cite{Sau81,Sau83} with the state of the art 
realistic NN potentials~\cite{PaXX}. Simultaneously a vigorous experimental
program was developed at Saclay. Thanks to the availability of an $^3$He and a
Tritium targets, the isoscalar and isovector components of the charge and the
magnetic elastic form factors of the three nucleon systems were
determined~\cite{Fro91}. They remarkably agreed with the predictions of the
Hanover group. The key to this success was not only the ``tour de force" in
operating the radioactive target, which was only made possible by an environment
of high technology, but also the close collaboration between a theoretical and
an experimental team.

I met Peter at this occasion and, starting from his wave functions, I extended
my diagrammatic approach~\cite{La81} to the disintegration channels of the three
nucleon systems~\cite{La91}. An experimental program was ongoing not only at 
Saclay but also at NIKHEF and Bates, in order to determine the wave function of
the few body systems.  Again, the experiments confirmed that Peter's wave
functions were doing a good job up to nucleon momentum of 500 MeV/c.

However, it became evident that the energy and the duty factor of the available
electron accelerators were not sufficient to map out the wave function at higher
momentum. Kinematical limits prevented to increase significantly the momentum
transfer. Rescattering and interaction effects were not negligible at such a low
energy. Finally the low duty factor prevented to single out the very small cross
sections, associated with high momentum transfers, from background.  

One of the reasons to built CEBAF at Jefferson Laboratory (JLab) was 
to overcome these drawbacks. The energy was increased by one order of magnitude 
(600 MeV to 6 GeV) and the duty factor by two orders of magnitude 
(1\% to 100\%). About seven years of operation have produced a few jewels 
in the disintegration of few body systems.

This meeting in honor of P. Sauer provides me with a good opportunity to
revisit the field in view of recent achievements and to evaluate future
developments.

\section{The Two Body Disintegration Channels}

The primary goal of the study of the (e,e'p) reaction on nuclei was, and still
is, the determination of the high momentum components of the nuclear wave
function, in view of the study of short range correlations and possible exotic
configurations. 

\subsection{The Low Energy Regime}

In the past the spectral functions measured at Saclay or Amsterdam suffered for
large corrections (about a factor two or more) due to Final State Interactions
(FSI) and Meson Exchange Currents (MEC). A survey of the  state of the art at
that time can be found in ref.~\cite{La91}. The corresponding experiments were
performed at low values ($\sim 0.3$~GeV$^2$) of the virtuality $Q^2$ of the
exchanged photon. 

\begin{figure}[hbt]
\begin{center}
\epsfig{file=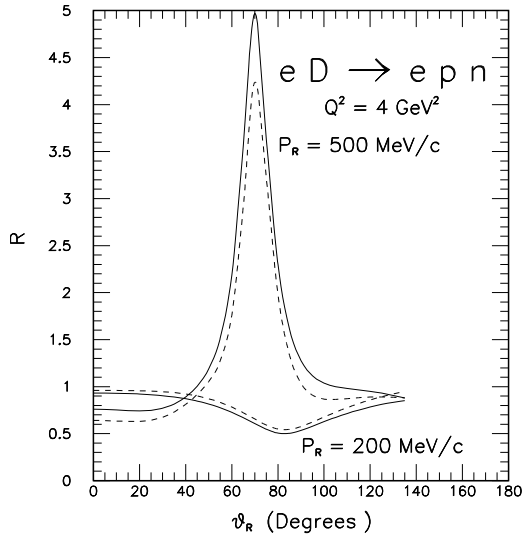,width=7cm}
\caption[]{The ratio between the full cross section and the contribution of the
quasi-free scattering.}
\label{sing}
\end{center}
\end{figure}

When CEBAF was decided, a common belief was that increasing $Q^2$ was the way to
suppress FSI and MEC contributions. This was partly true, since both the FSI and
MEC amplitudes involve a loop integral, which connects the nuclear bound and
scattering states and which is expected  to decrease when $Q^2$ increases as
form factors do. But this was partly wrong, since the singular part of the FSI
integral does not depends on $Q^2$, besides the trivial momentum dependency of
the elementary operators. It comes from unitarity, and corresponds to the
propagation of an on-shell nucleon. It involves on-shell elementary matrix
elements and it is maximum when the kinematics allows for rescattering on a
nucleon at rest. This happens in the quasi-free kinematics, when $X=Q^2/2m\nu=1$ (being
$\nu$ the energy of the virtual photon).

Fig.~\ref{sing} exhibits these features. It shows the angular distribution,
against the neutron angle $\theta_R$, of the ratio between the full cross 
section of the
D(e,e'p)n reaction and the quasi free contribution, when the momentum $P_R$ of
the recoiling neutron is kept constant. FSI are maximum near $\theta_R=
70^{\circ}$ where $X=1$ and on-shell rescattering is maximized. At low values of
the recoil momentum ($P_R= 200$~MeV/c), on-shell nucleon rescattering reduces
the quasi free contribution, as expected from unitarity (a part of the strength 
of the quasi elastic channel is transferred to inelastic ones).  At high values
of the recoil momentum ($P_R= 400$~MeV/c) the quasi free contribution strongly
decreases as the nucleon momentum distribution: on-shell rescattering takes
over and dominates.  

\begin{figure}[hbt]
\begin{center}
\epsfig{file=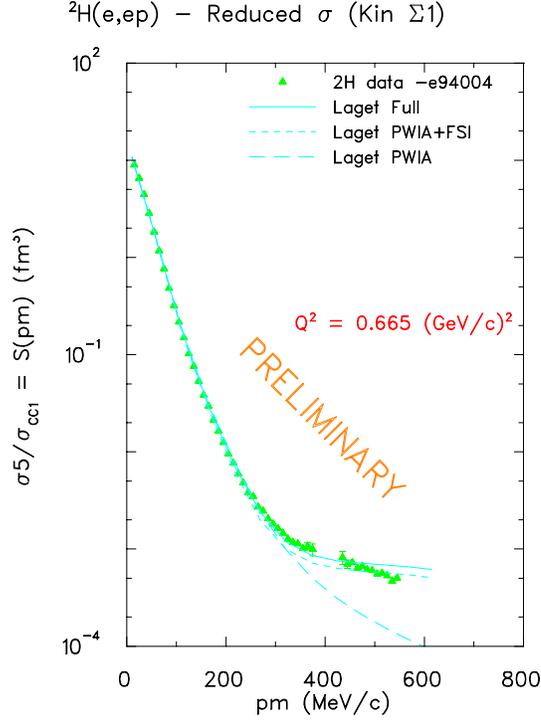,width=7cm}
\caption[]{The momentum distribution in the D(e,e'p)n reaction at X=1.}
\label{Ulmer}
\end{center}
\end{figure}

An experiment~\cite{Ulm02} recently performed at JLab confirms this behavior.
Fig.~\ref{Ulmer} shows the recoil momentum distribution up to 600 MeV/c, in the
quasi elastic kinematic (X=1). Above $P_R=400$~MeV/c, FSI (dashed line) 
dominate while MEC (full line) contribute to a lesser extent. 

Although the virtuality ($Q^2= 0.665$~GeV$^2$) of the exchanged photon is about
twice as much than what was achieved twenty years ago, the relative kinetic
energy of the two outgoing nucleons ($T_L= Q^2/2m= 337$~MeV) is still low enough
to rely on the partial wave expansion of the np scattering amplitude. The curves
are the results of the  original model~\cite{La87}, where both the on-shell and
half off-shell np scattering amplitudes are solution of the Lippman-Schwinger
equation with the same potential (Paris) as for the bound state. 

\subsection{The High Energy Regime}

At higher energies (let's say when the relative kinetic energy of the outgiong
fragment exceeds 500~MeV or so), too many partial waves enter into the game and
their growing inelasticities prevent to compute the scattering amplitude from a
potential. It is better to use a global parameterization of the NN scattering
amplitude. On general ground, it can be expanded as follows
\begin{equation}
T_{NN} = \alpha + i\gamma (\vec{\sigma_1}+\vec{\sigma_2})\cdot \vec{n}
+\;\mathrm{spin-spin} \;\mathrm{terms} 
\end{equation}

Above 500 MeV, the central part $\alpha$ dominates and is almost entirely
absorptive, and takes the simple form
\begin{equation}
\alpha = \frac{k}{4\pi} \;(\epsilon + i)\;\sigma_{NN} \;\exp[\frac{b}{2}t] 
\end{equation}
In the forward direction its imaginary part is related to the total
cross section $\sigma_{NN}$, while the slope parameter $b$ is related to the
angular distribution of NN scattering. Both can be determined from the
experiments performed at Los Alamos, Saturne and Cosy. The ratio $\epsilon$,
between the real and imaginary part of the amplitude, is small and does not
exceed 0.2. The spin-orbit and spin-spin terms are related to polarization
observables, but I have not yet implemented them in the three body codes.

Such a parameterization is very  convenient to compute the rescattering 
amplitude, but leads only to an accurate prediction of its singular part
(on-shell scattering). Contrary to low energy, there is unfortunately  no way 
to constrain the half-off part of the NN scattering amplitude, and one can get
only estimate of the principal part of the rescattering amplitude. It turns out
that it does not dominate at high energy. So, the method is founded on solid
grounds in the quasi-elastic kinematics (X$\sim$1). Away, it tells us in which
kinematics FSI are minimized. 

\begin{figure}[hbt]
\begin{center}
\epsfig{file=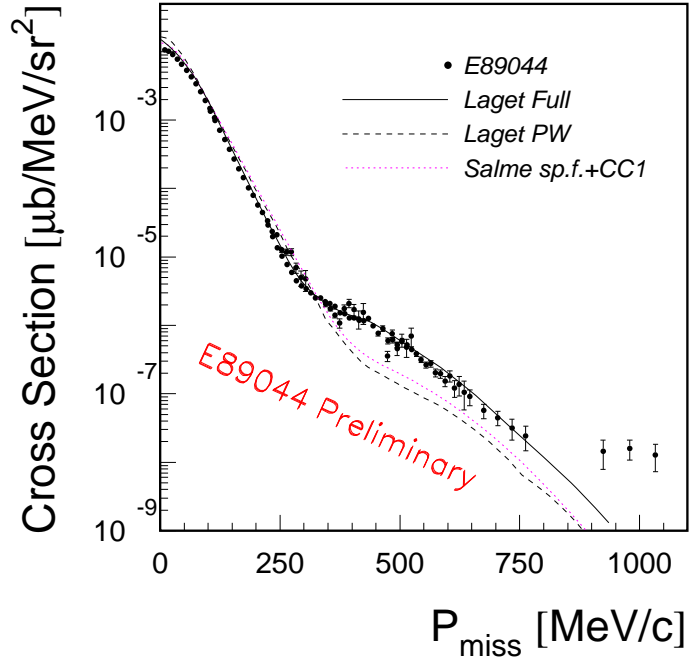,width=9cm}
\caption[]{The momentum distribution in the $^3$He(e,e'p)d reaction at X=1 and
$Q^2=1.55$~GeV$^2$.}
\label{he3pd}
\end{center}
\end{figure}

Fig.~\ref{he3pd} shows how well this method reproduces the cross section of the
$^3He(e,e'p)d$ reaction recently measured at JLab~\cite{Hig02} at $Q^2=1.55$~GeV$^2$, in the
quasi-free kinematics (X=1). At such a high virtuality, the relative kinetic
energy between the outgoing proton and deuteron is $T_L= 850$~MeV, where the NN
cross section reaches its maximun and becomes flat around $\sigma_{NN}= 47$mb.
Again, FSI reduces the quasi-free contribution below 300~MeV/c and overwhelms it
by more than a factor five around 500~MeV/c. Details on the model are given in
ref.~\cite{Lag94}.

To summarize, a fair agreement with the recent JLab data has been reached in the
quasi-elastic regime, up to recoil momentum of the order of 1 GeV/c, provided
that the NN scattering amplitude relevant to the same energy range is used. In
order to determine the high momentum components of the nuclear wave function,
one has to go away the quasi-elastic kinematics: as demonstrated in 
Fig.~\ref{sing}, this occurs  in parallel or anti-parallel kinematics, where 
on-shell nucleon rescattering is suppressed.

\section{The Three Body Disintegration Channels}

Two body short range correlations are the primary source of high momentum
components in the nuclear wave function. Since they are strongly coupled to high
energy state in the continuum, two nucleon production experiments are the
natural way to reveal them. However, above the pion production threshold, one 
must  eliminate meson production channels and perform exclusive experiments. 
CLAS (CEBAF Large Acceptance Spectrometer) at JLab is the ideal detector for
disentangling and studying multiparticle channels. 

However, CLAS is not an hermetic detector: there is a hole in the forward 
direction
as well as in the backward direction; also the coils, which provide the toroidal
electromagnetic field,  induce dead zones in the acceptance. So the only fair
way to compare the data to any theory is to use the same cuts. To that end, G.
Audit and I, we have developed a Monte Carlo programme which simulates the
$^3$He(e,e'2p)n and $^3$He($\gamma$,2p)n reactions within the CLAS acceptance. 
The event generator is the code which I developed twenty years ago and which I
updated to take into account  the high energy parameterization of the NN
scattering amplitude, as well as relativistic effects.

Let me discuss  first the real and then the virtual photon sectors.

\subsection{Real Photons}

\begin{figure}[hbt]
\begin{center}
\epsfig{file=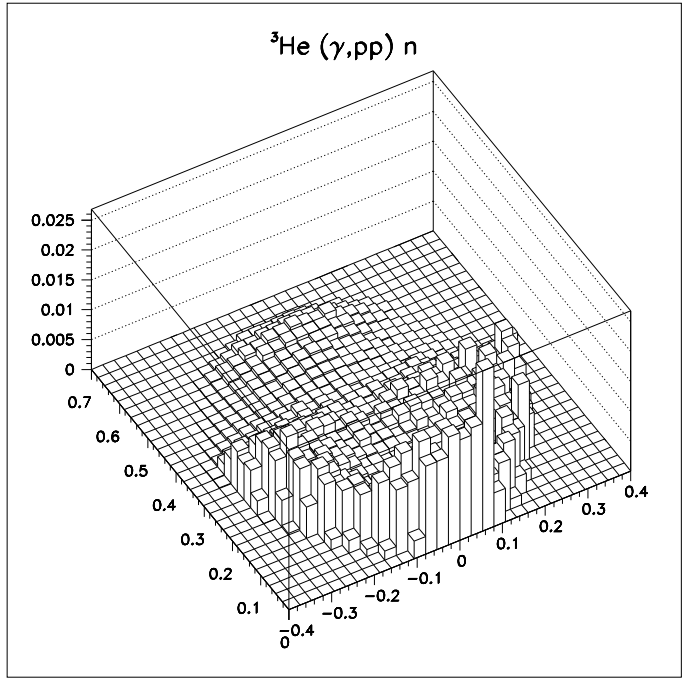,width=9cm}
\caption[]{The Dalitz plot of the $^3$He($\gamma$,2p)n reaction at 550 MeV. The
bottom-right axis is proportional to the difference $T_1-T_2$ between the
kinetic energy of the two protons, while the left axis is proportional to the
kinetic energy of the neutron $T_n$.}
\label{dalitz}
\end{center}
\end{figure}

Fig.~\ref{dalitz} shows the result of such a simulation within the CLAS
acceptance. The model~\cite{La88} has been calibrated against previous data 
recorded below 800 MeV in restricted
parts of the phase space: two magnetic spectrometers~\cite{Au89,Au93} in the two
body disintegration sector (neutron at rest); two magnetic
spectrometers~\cite{Au91} and DAPHNE detector~\cite{Au97} in the three body
disintegration sector.  

CLAS results~\cite{Nic02} enlarge the available data set, not only by covering
the full phase space but also by extending the energy range in a single shot.
Cuts can be made in various places of the Dalitz plot, in order to emphazis
different mechanisms and access different aspects of the three body dynamics.
For instance, the peak at bottom of the Dalitz plot (small $T_n$) corresponds to
the disintegration of a pair of protons at rest. Since this pair has no dipole
moment, MEC associated with the $\Delta$ resonance are suppressed and one can
probe higher order MEC. The peaks on the left and right edges of the Dalitz plot
correspond to rescattering within a proton neutron pair, while the other proton
takes almost all the available energy. Finally, in the center of the Dalitz plot
the energy is shared by the three nucleons: this is the place where three nucleon
effects dominate. Preliminary results show a clear distinction between two and
three body effects and follow the trends of the model. I refer to
ref.~\cite{Nic02} for a more detailed account of the preliminary CLAS data.

\subsection{Virtual Photons}

The $^3$He(e,e'2p)n reaction was advocated~\cite{La87} as the best way to access
two proton correlations in few body systems. In particular, in kinematics where
the neutron is almost at rest, the transverse part of the amplitude is
suppressed since the pair of proton has no dipole moment. It also turns out that
three body mechanisms are suppressed, since they prefer kinematics where the
momentum transfer is shared by the three nucleon. On the contrary, the coupling
of a longitudinal photon the the proton pair is not suppressed, and the only
sizeable correction is due to FSI between the two outgoing protons.

\begin{figure}[hbt]
\begin{center}
\epsfig{file=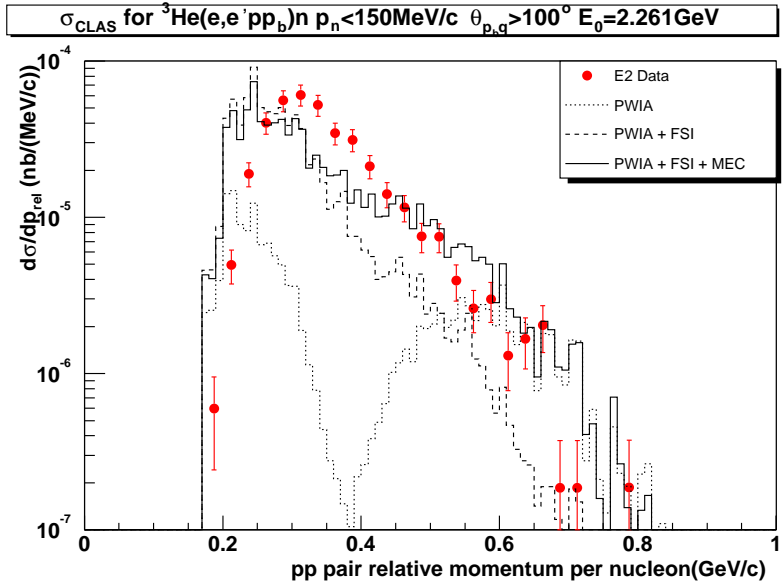,width=11.5cm}
\caption[]{The cross section of the $^3$He(e,e'2p)n reaction, when the neutron
is almost at rest (CLAS preliminary). The fast proton is emitted forward, while
 the slow proton is
emitted backward. The dotted line corresponds to one body processes. The dashed
line includes FSI. The full line includes also MEC.}
\label{back}
\end{center}
\end{figure}

CLAS provides us with the first comprehensive study of this 
channel~\cite{Zha02,Gi01}. Fig.~\ref{back} shows the momentum distribution of 
the pair of protons almost at rest in $^3$He. The plane wave calculation 
exhibits
the characteristic node of the S-wave part of its wave function, which is the
only one which survives when the pair is at rest. Of course, FSI fill in this
hole (nature does not like holes!), and MEC contribute moderatly at high
momentum. Both the theory and the experiment are integrated within the same cuts
(for instance, the fall off of the cross section below 0.2 GeV$^2$ reflects the 
CLAS acceptance) and there is no normalization factor (absolute comparison). 
The wave function of P. Sauer does a good job in a configuration where it has
never been checked before.

Increasing the virtuality from $Q^2\sim 1$~GeV$^2$ up to a few GeV$^2$ 
will provide us with a more stringent test. Since this kinematic is far from 
the 
quasi-elastic one, on-shell nucleon rescattering is not dominant, and FSI and
MEC will be suppressed. 

\begin{figure}[hbt]
\begin{center}
\epsfig{file=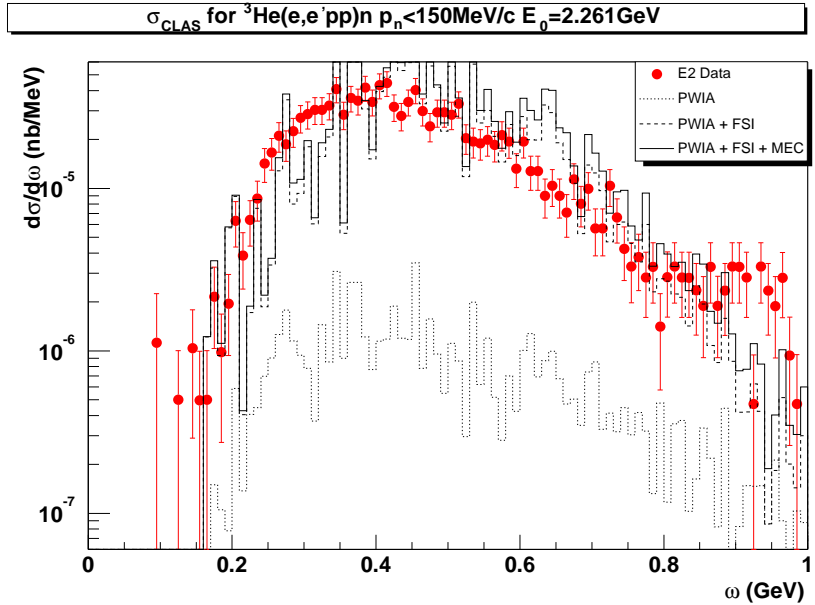,width=8cm,angle=-90}
\caption[]{The cross section of the $^3$He(e,e'2p)n reaction, when the neutron
is almost at rest(CLAS preliminary). The two protons are emitted in a symmetric
 way against the 
virtual photon direction. The dotted line corresponds to one body processes.
The dashed line includes FSI. The full line includes also MEC.}
\label{sym}
\end{center}
\end{figure}

On the contrary, on-shell nucleon rescattering dominates the cross section in
the kinematics reported in Fig.~\ref{sym}. Here the two protons are required to
be emitted in a symmetric way around the direction of the virtual photon, and
the cross section is plotted against the energy of the virtual photon. The good
agreement between the theory and the experiment tells us that our description of
the on-shell rescattering matrix element is fine over a wide energy range
($0.2<T_L<2$~GeV).

Other cuts have been applied to CLAS data. For instance, one can select a very
fast neutron and look for a proton pair almost at rest in $^3$He. In a plane
wave picture, this gives access to the relative wave function of the two
protons. However, in the interesting situation where the relative momentum
between the two protons is large, the three nucleons share the total energy and
momentum and three body mechanisms contribute significantly. I refer to
ref.~\cite{Wei01} for a more detailed accounts of such a study.

\section{Perspectives}

The operation of CEBAF at Jefferson Laboratory, over the past few years, has
confirmed the expectations which  we had twenty years ago, but has enlarged
in a considerable way the data set, both in accuracy and in energy as well 
as momentum range.

In the real photon sector, the three body photo disintegration of $^3$He is
dominated by three body mechanisms related to the on-shell propagation of a pion
between the three nucleons. The extension to the virtual photon sector gives
access to three body Meson Exchange Currents, which may be related to three body
forces.

In the virtual photon sector, the two-body electrodisintegration of $^3$He, and
of a pair of two nucleons bound in $^3$He, opens windows on the momentum
distribution at short distance. However special care has to be taken in order to
avoid on-shell nucleon scattering which dominates the electrodisintegration
cross section in a well defined part of the phase space, namely near the quasi
free kinematics. The best candidate to study the short range components of the 
nuclear wave function is the parallel kinematics, where the outgoing nucleons are
emitted along the direction of the virtual photon. In that kinematics FSI and
MEC contributions are reasonably small and are expected to decrease when the
virtuality $Q^2$ increases in the range of several GeV$^2$.

Alternately, one may take advantage of the strong on-shell scattering
contribution to study the rescattering of exotic components of the nucleon (and
other hadrons) wave function. The study of Color Transparency (CT) is an example
of such a possibility. CT is a natural consequence of QCD. When a photon of
high virtuality $Q^2$ interacts elastically with a nucleon, it selects
configurations with small transverse extension which undergo less rescattering
when they travel in the nuclear medium. However it turns out that in the range
of virtuality currently available ($Q^2<6$~GeV$^2$) the lifetime of such small
configurations is comparable to the internucleon distance and much smaller than
the radius of a nucleus. This explains why no signal of Color Transparency has
been reported so far. A better way would be to work in kinematics where
rescatterings between to nucleons are maximized, such as in Fig.~\ref{sing}. At
low virtuality $Q^2$ the hight and the width of the rescattering peak are on
solid grounds, since the matrix element depends of on-shell elementary
amplitudes (which can be borrowed from the corresponding elementary channels)
and on the low momentum part of the nuclear wave function. At higher virtuality,
CT will cause the peak to decrease and  its width to increase. I refer to my
contribution in ref.~\cite{La00} for a more detailed discussion.

While the present energy of CEBAF (6 GeV) allows to test this conjecture at low
virtually ($Q^2<6$~GeV$^2$), only its energy upgrade to 12 GeV will allows to
reach high enough virtuality ($Q^2\sim 12$~GeV$^2$) to see the effects of color
transparency.

\begin{acknowledge}
This work was supported in part by the European Commission under Contract
HPRN-CT-2000-00130.
\end{acknowledge}

\end{document}